\documentclass[sigconf]{acmart}

\usepackage{booktabs}
\usepackage{graphicx}
\usepackage{subcaption}
\usepackage{amsmath}
\usepackage{algorithm}
\usepackage{algpseudocode}

\renewcommand\footnotetextcopyrightpermission[1]{} 
\usepackage{subcaption}
\usepackage{svg}
\usepackage{booktabs}
\usepackage{multirow}
\usepackage{makecell}
\usepackage{xspace}
\usepackage{enumitem}
\usepackage{xcolor}

\newcommand{\sysname}{\textsf{UniExit}\xspace}

\newcommand{\eg}{\emph{e.g.},\xspace}

\newcommand{\rev}[1]{#1}

\acmConference[SenSys '27]{The International Conference on Embedded Artificial Intelligence and Sensing Systems}{2027}{Boulder, CO, USA}

\author{Yuhang Guan}
\affiliation{%
  \institution{The Hong Kong University of Science and Technology (Guangzhou)}
  \city{Guangzhou}
  \country{China}
}
\email{asterism178@163.com}

\author{Huanle Zhang}
\affiliation{%
  \institution{Shandong University}
  \city{Qingdao}
  \country{China}
}
\email{dtczhang@sdu.edu.cn}

\author{Xiaomin Ouyang}
\affiliation{%
  \institution{The Hong Kong University of Science and Technology}
  \city{Hong Kong}
  \country{China}
}
\email{xmouyang@cse.ust.hk}

\author{Zhidan Liu}
\authornote{Corresponding author.}
\affiliation{%
  \institution{The Hong Kong University of Science and Technology (Guangzhou)}
  \city{Guangzhou}
  \country{China}
}
\email{zhidanliu@hkust-gz.edu.cn}

\begin{document}

\title{\sysname: A Unified Latency-Aware Early-Exit Runtime for Heterogeneous Edge AI}

\begin{abstract}
Edge applications increasingly combine CNN-based perception with edge-deployable Transformer-based language models, but deploying such heterogeneous workloads on resource-constrained devices is challenging because inference latency varies with both input difficulty and runtime hardware state. Early-exit inference reduces computation by terminating at intermediate stages, yet existing methods are often model-specific and rely on static latency profiles that become stale under dynamic edge conditions. This paper presents \sysname, a unified latency-aware early-exit runtime for heterogeneous edge AI. \rev{\sysname abstracts CNN blocks and Transformer layers as ordered stages and applies a common latency-aware routing policy while retaining model-specific quality mechanisms. It predicts the cost of reaching the next exit from runtime state and inference progress to make budget-aware continue-or-exit decisions.} \sysname combines runtime-state-aware latency prediction, cache-based prediction reuse, trend-aware fallback, and CNN-specific exit stabilization to provide low-overhead and robust online control. We implement \sysname on multiple edge and edge--cloud platforms and evaluate it with representative CNN and Transformer-based language-model workloads. Across single-workload, concurrent heterogeneous, and edge--cloud settings, \sysname consistently improves deadline satisfaction and latency--accuracy trade-offs over existing progressive inference baselines, while reducing redundant exit-module computation and task-level energy consumption.
\end{abstract}

\maketitle

\begin{figure*}[!t]
    \centering
    \includegraphics[width=0.90\linewidth]{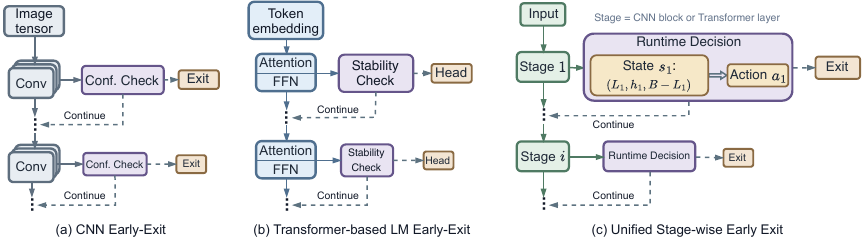}
    \caption{\rev{From model-specific early exits to a shared stage-wise runtime view. (a) CNN early-exit methods use branch classifiers and confidence checks. (b) Transformer-based language-model methods use intermediate heads and stability checks. (c) \sysname abstracts both as staged runtime decisions driven by execution progress, runtime state, and latency budget, without using model-specific confidence or stability signals for routing.}}
    \label{fig:early_exit_edge_ai}
\end{figure*}

\section{Introduction} \label{sec:introduction}

With advances in edge computing and on-device hardware, AI inference is increasingly moving from cloud servers to edge devices, closer to users, sensors, and data sources~\cite{edge_intelligence_survey, on_device_ai_survey, edge_review}. As edge applications become more intelligent and interactive, they are evolving from single-model deployments to heterogeneous AI workloads~\cite{pantheon, edgeserving} that combine CNN-based visual perception~\cite{e4} with edge-deployable Transformer-based language models for language understanding and interaction~\cite{edge_llm_survey, mobile_edge_llm_survey}. These workloads differ in task semantics, computation patterns, and latency--accuracy requirements, yet share a common deployment need: \textit{delivering timely and reliable inference under resource-constrained and dynamic edge environments}. Representative scenarios include in-vehicle perception~\cite{autonomous_driving_survey}, interactive AI assistants~\cite{edge_llm_survey, mobile_edge_llm_survey}, augmented reality~\cite{edge_ar_mobicom, ar_human_city_survey}, and context-aware mobile sensing~\cite{context_aware_edge_ai, mobile_sensing_deep_learning}, where latency directly affects responsiveness, user experience, and reliability.

A key opportunity in latency-critical deployments is that \textit{not all inputs require the same amount of computation}. Some visual inputs can be recognized from shallow features~\cite{branchynet}, while others require deeper semantic representations~\cite{msdnet, shallowdeep}; similarly, some language-model tokens can be predicted from intermediate layers~\cite{deebert, layerskip}, while others require deeper contextual refinement~\cite{specee}. Full-model inference ignores this variability and executes the entire model for every input or token, wasting computation on easy cases. Progressive inference~\cite{early_exit_survey, e4, distrEE}, or early-exit inference, addresses this inefficiency by inserting candidate exits into a neural network and terminating inference once an intermediate prediction is reliable. Compared with model pruning~\cite{pruning_quantization_survey} and computation offloading~\cite{neurosurgeon, edgent, jointopt}, early-exit inference provides a flexible on-device mechanism for runtime adaptation while preserving the option of deeper inference when needed.

Existing early-exit techniques have been studied for both CNNs~\cite{branchynet, hmpsoc, ceed} and Transformer-based language models~\cite{layerskip, specee}. CNN methods typically attach branch classifiers to intermediate convolutional or residual blocks~\cite{branchynet}, while language-model methods use intermediate prediction heads or layer-wise confidence signals to decide whether later Transformer layers can be skipped~\cite{deebert, pabee, layerskip}. These designs demonstrate the potential of adaptive computation, but they are usually developed for specific model families and tightly couple confidence estimation with \textbf{model-specific exit policies}. As a result, existing early-exit systems provide limited support for managing heterogeneous edge AI workloads under a unified runtime objective, even though CNN and language-model workloads increasingly coexist on the same edge platforms.

In this paper, we make a key observation: \textit{CNNs and Transformer-based language models expose a common staged decision structure during early-exit inference}, despite their different task semantics and computation patterns. CNNs execute a sequence of convolutional or residual blocks, while Transformer-based language models execute a sequence of Transformer layers. \rev{At selected stages, both expose candidate outputs that can be returned upon termination. Early-exit inference can therefore be viewed as a runtime-control problem driven by execution progress, predicted remaining cost, runtime state, and latency constraints, while keeping model-specific exit and quality mechanisms unchanged.}

However, making such runtime decisions reliable on edge devices is challenging because inference latency is not static. Many early-exit systems~\cite{spinn, e3, densepipe} rely on offline profiling to construct static latency tables for different exit paths and then use these tables to estimate online execution cost~\cite{edgeml, eenet, e4}. This design assumes that segment latency remains stable during deployment. In practice, edge devices dynamically adjust CPU, GPU, and memory-controller states~\cite{dvfo} to balance performance, power, and temperature~\cite{dvfs_aware_dnn, e4, jetson_characterization}, and may also experience resource contention from co-running workloads~\cite{pantheon, rt_mdl, edgeserving}. Consequently, the same model segment can exhibit different latency across time. As a result, \textbf{static latency profiles can become stale}, leading to unnecessary exit execution, premature exits, or deadline violations.

To address these challenges, we propose \sysname, a \textbf{unified latency-aware early-exit runtime} for dynamic edge environments. \rev{\sysname provides a common stage-wise interface and shared latency-aware routing for CNN and language-model workloads, while retaining model-specific exit and quality mechanisms.} Rather than relying solely on static latency profiles, \sysname incorporates runtime state and inference progress into exit decisions. At each candidate exit, \sysname estimates whether continuing to the next stage can still satisfy the latency budget. If the predicted cost fits within the remaining budget, \sysname bypasses the current exit module and continues deeper inference; otherwise, it executes the current exit and returns an early result. \rev{In this way, \sysname selects the exit position according to real-time latency feasibility, independently of model-specific quality signals.}

\sysname makes this runtime practical through three lightweight mechanisms. First, it trains \textbf{runtime-state-aware latency predictors} that estimate stage and exit-module latency using hardware frequencies, resource utilization, and computation-unit identifiers. Second, it uses \textbf{cache-based prediction reuse} to avoid redundant predictor invocations under recurring runtime states, reducing online decision overhead. Third, it incorporates \textbf{trend-aware fallback} to make routing more conservative when abrupt hardware-state changes may invalidate recent predictions. Beyond the shared runtime, \sysname also includes a CNN-specific offline stabilization objective that reduces confidence degradation and prediction flips across CNN exits, while preserving the same online interface used for language-model workloads.

We implement \sysname on multiple edge and edge--cloud platforms and evaluate it with representative CNN and Transformer-based language-model workloads. Our evaluation covers isolated single-workload inference under static and dynamic hardware states, concurrent heterogeneous workloads, runtime overhead, ablation studies, and CNN-based edge--cloud split inference. Across these settings, \sysname consistently improves deadline satisfaction and latency--accuracy trade-offs over existing progressive inference baselines. It achieves over 95\% deadline satisfaction in single-workload and edge--cloud settings, remains robust under concurrent heterogeneous workloads, and reduces redundant exit-module computation and task-level energy consumption.

This paper makes the following contributions:
\begin{itemize}[leftmargin=1.2em,itemsep=0.2em,topsep=0.2em]
    \item \rev{We formulate CNN and Transformer-based language-model early-exit inference as a shared stage-wise runtime-control problem, with common latency-aware routing and model-specific quality mechanisms. This abstraction improves runtime code reuse and reduces duplicated logic and integration complexity.}

    \item We design \sysname, a latency-aware early-exit runtime that incorporates runtime state, inference progress, and latency budgets to make adaptive exit decisions under dynamic edge environments.

    \item We make the runtime practical through lightweight latency prediction, cache-based reuse, fallback routing under volatile hardware states, and CNN-specific exit stabilization.
        
    \item We implement \sysname on multiple edge and edge--cloud platforms and evaluate it across single-workload, concurrent heterogeneous, and edge--cloud settings, showing that \sysname delivers consistently stronger deadline-aware inference across deployment settings, with better latency--accuracy trade-offs and lower computation and energy overhead.
\end{itemize}

\section{Motivation} \label{sec:motivation}

\subsection{Early-Exit Inference in Edge AI}

In latency-critical edge inference, \textit{not all inputs or tokens require the same amount of computation}~\cite{early_exit_survey, e4, edgeserving}. Some images can be recognized from shallow visual features, while harder ones require deeper semantic representations~\cite{branchynet,msdnet,shallowdeep}. Likewise, some language-model tokens can be predicted from intermediate Transformer layers, while others need deeper contextual refinement~\cite{pabee, layerskip, specee}. Full-model execution for every input or token therefore wastes computation on easy cases.

Early-exit inference reduces redundant computation through input-dependent conditional execution~\cite{early_exit_survey, e4, hmpsoc}. Instead of executing the full model, it inserts intermediate exits that return a prediction once the current representation is sufficiently reliable. As shown in Figure~\ref{fig:early_exit_edge_ai}(a), CNN early-exit networks attach branch classifiers to selected convolutional or residual blocks~\cite{branchynet, msdnet, shallowdeep}. Each branch produces a tentative prediction and confidence score from the current feature representation. If the confidence exceeds a threshold, the image exits early and skips the remaining blocks; otherwise, it continues deeper. This image-level branch execution saves computation for easy samples while retaining deeper paths for harder ones when the latency budget allows.

Figure~\ref{fig:early_exit_edge_ai}(b) illustrates a related early-exit pattern in Transformer-based language models, but with a different execution granularity~\cite{densepipe, layerskip, specee}. Since Transformers are organized as ordered layers, intermediate prediction heads or confidence/stability signals can evaluate the current hidden state after selected layers~\cite{deebert, pabee, layerskip}. During generation, the exit policy decides whether a token can be emitted from the current layer or should continue deeper for refinement. In latency-constrained deployments, this decision must also respect the remaining time budget, motivating the runtime abstraction introduced later. Unlike CNN classification, language-model early exit is often token-level, allowing easy and hard tokens within the same sequence to follow different computational depths.

\subsection{Limitations of Existing Early-Exit Systems}

Early-exit inference provides an effective mechanism for adaptive computation, but existing systems are often built around specific model families, exit modules, and latency assumptions~\cite{spinn, e3, seeb_gpu}. This design style creates a \textit{mismatch} with emerging edge AI scenarios, where CNN-based perception and Transformer-based language-model workloads may coexist on the same device or within the same edge--cloud pipeline. We identify two key limitations.

\textbf{Limitation 1: Exit control is model-specific.}
Existing methods usually couple confidence estimation, exit policy, and accuracy--latency tradeoff within a model-specific pipeline. CNN early-exit methods typically attach branch classifiers to intermediate convolutional or residual blocks and make image-level exit decisions based on branch confidence~\cite{branchynet, shallowdeep}, sometimes together with latency or deadline constraints~\cite{e4, densepipe, seeb_gpu}. In contrast, Transformer-based language-model methods use layer-wise heads, token-level stability, or intermediate logits to decide whether the current token should exit or continue to deeper layers~\cite{deebert, layerskip, specee}. Although both pipelines follow an \textit{exit-or-continue} pattern, their control logic is embedded in different model structures and execution granularities. This makes it difficult to reuse exit policies across workloads or expose a shared runtime controller for heterogeneous edge inference.

\textbf{Limitation 2: Latency awareness relies on static profiles.}
Even when latency constraints are considered, many early-exit systems estimate execution cost using offline latency profiles. A typical system profiles each block, layer, exit module, or remaining path under a fixed hardware state, then uses the latency table to determine whether continuing to a deeper exit is affordable under the remaining deadline~\cite{spinn, e3, seeb_gpu}. This approach is simple and efficient, but it implicitly assumes that profiled latency remains valid during deployment. This assumption is fragile on edge devices, where dynamic voltage and frequency scaling (DVFS), thermal effects, background workloads, memory-frequency changes, and resource contention can change execution latency over time~\cite{dvfs_robust, malva, nnperf, miriam}. Once the profiled latency becomes stale, the remaining-latency estimate can mislead the exit policy: an optimistic estimate may continue too deeply and miss the deadline, while a conservative estimate may trigger premature exits and reduce prediction quality.

\begin{figure}[t]
    \centering
    \includegraphics[width=0.75\linewidth]{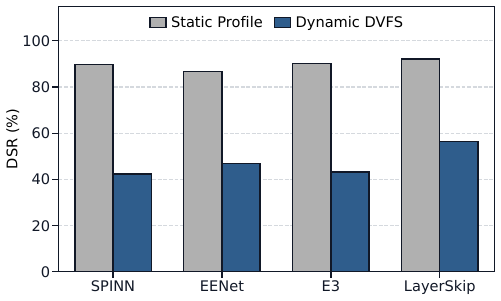}
    \caption{CNN and language-model early-exit methods suffer substantial DSR drops under dynamic hardware states.}
    \label{fig:dsr_static_dynamic}
\end{figure}

To quantify the impact of stale latency profiles, we conduct a preliminary study with representative early-exit methods for both CNN and Transformer-based language-model workloads. Specifically, we evaluate SPINN~\cite{spinn}, EENet~\cite{eenet}, and E3~\cite{e3} as CNN early-exit methods, and LayerSkip~\cite{layerskip} as a Transformer-based language-model early-exit method on an edge-deployable small language model; detailed experimental settings are described in \S\ref{subsec:expsetup}. This preliminary study focuses on DVFS-induced hardware variability. As shown in Figure~\ref{fig:dsr_static_dynamic}, all methods achieve high deadline satisfaction rates (DSRs) under static hardware conditions, but their DSRs drop substantially under dynamic DVFS conditions. The degradation appears in both CNN and language-model workloads, indicating that the problem is not tied to a particular model family. These results show that static latency profiles can support effective exit decisions in controlled environments, but become unreliable when hardware performance changes dynamically at runtime~\cite{dids}.

These limitations suggest two requirements for edge early-exit inference. First, \textit{exit decisions should be online latency-aware}, accounting for current execution progress, remaining latency budget, and runtime hardware state instead of relying solely on static profiles. Second, \textit{such control should be reusable across model families}, rather than redesigned separately for CNN and language-model workloads. These requirements motivate the unified stage-wise view introduced next.

\subsection{A Unified Stage-Wise View}

\rev{CNN and Transformer-based language-model early-exit systems differ in model structure and exit semantics, but both execute through sequential model blocks. We define a stage as a model block with strong internal data dependencies and use its boundary as a runtime decision point. A stage can correspond to a residual block in CNNs or a Transformer layer in language models. Such boundaries provide a consistent decision granularity while avoiding runtime control inside asynchronous kernels or overlapping execution regions that could disrupt execution concurrency. At each candidate exit, the current output is returned if continuing would exceed the remaining latency budget.}

We therefore abstract early-exit inference as a \textit{staged runtime decision process}. For each inference request, $B$ denotes the application-specified latency budget, or relative deadline. As illustrated in Figure~\ref{fig:early_exit_edge_ai}(c), at stage $i$, the runtime observes
\begin{equation}
\rev{s_i = (L_i, h_i, B - L_i),}
\end{equation}
\rev{where $L_i$, $h_i$, and $B-L_i$ denote elapsed latency, runtime state, and remaining latency budget, respectively.} The runtime then chooses
\begin{equation}
a_i \in \{\textit{exit}, \textit{continue}\}.
\end{equation}
\rev{On exit, the system executes the current exit module and returns $y_i$; otherwise, it proceeds to stage $i+1$. Exit modules and outputs remain model-specific, while runtime monitoring, latency estimation, budget tracking, and routing are shared across workloads.}

This abstraction directly addresses the two limitations above. \rev{It provides a common interface for model-specific exits and makes routing aware of execution progress, remaining budget, and hardware state rather than static latency profiles alone.} This unified stage-wise view forms the basis for \sysname, a latency-aware early-exit runtime for heterogeneous edge AI workloads.

\section{System Overview} \label{sec:overview}

\begin{figure}[!t]
    \centering
    \includegraphics[width=1.0\linewidth]{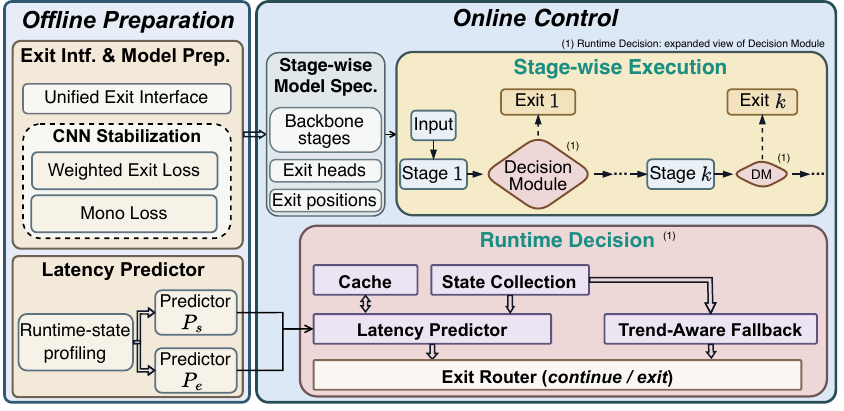}
    \caption{Overview of \sysname. Offline preparation builds exit interfaces, stabilization, and latency predictors; online execution invokes a decision module before candidate exits. The Runtime Decision block expands this module by monitoring runtime state, predicting next-exit cost, and routing each decision to continue or exit.}
    \label{fig:uniexit_overview}
\end{figure}

\subsection{Design Goals} \label{subsec:goals}

Motivated by the limitations and requirements identified above, the design of \sysname targets four design goals:
\begin{itemize}[leftmargin=1.2em,itemsep=0.2em,topsep=0.2em]
    \item \rev{\textbf{Unified control.} Provide CNN and language-model workloads with a shared stage-wise latency-aware routing interface while preserving model-specific quality mechanisms.}

    \item \textbf{Online latency awareness.} Make exit-or-continue decisions using elapsed latency, runtime state, and the remaining latency budget, rather than relying only on static latency profiles.

    \item \textbf{Low overhead.} Keep latency prediction and exit routing lightweight enough that online control does not offset the computation saved by early exits.

    \item \textbf{Runtime robustness.} Maintain deadline satisfaction under dynamic hardware states and workload interference, avoiding both overly optimistic continues that miss deadlines and overly conservative exits that reduce prediction quality.
\end{itemize}

\subsection{\sysname Architecture} \label{subsec:architecture}

Figure~\ref{fig:uniexit_overview} shows the architecture of \sysname. To realize the goals above, \sysname separates model-specific exit preparation from shared runtime control. In the offline phase, \sysname prepares exit modules and runtime-state-aware latency predictors for different workloads. For CNN workloads, \sysname trains early-exit branches with CNN-specific stabilization to improve exit reliability; for Transformer-based language models, it reuses layer-wise exit heads or pretrained early-exit models. In parallel, \sysname collects latency measurements under representative runtime states and trains compact predictors for estimating stage and exit-module latency.

During online inference, \sysname executes the model as ordered stages and invokes a lightweight decision module before each candidate exit. The lower Runtime Decision block in Figure~\ref{fig:uniexit_overview} expands the internal logic of this module. It uses the elapsed latency $L_k$, the predicted cost of reaching the next candidate exit $\hat{\Delta L}_{k\rightarrow k+1}$, the current runtime state, and the latency budget $B$. \sysname continues to deeper computation when $L_k + \hat{\Delta L}_{k\rightarrow k+1} + \Delta S \le B$, and exits at the current stage otherwise, where $\Delta S$ is a safety margin for volatile runtime conditions. This decision placement avoids unnecessary exit-module execution when the remaining budget still allows deeper inference.

To keep the runtime lightweight and robust, \sysname integrates cache-based prediction reuse and trend-aware fallback. The cache reduces repeated predictor invocations under recurring runtime states, while fallback makes routing more conservative when runtime states become volatile. Together, they enable robust, low-overhead latency-aware control across heterogeneous edge workloads. The following sections describe these components in detail.

\section{Offline Support for Runtime Control}
\label{sec:offline}

\begin{table}[!b]
    \centering
    \caption{Offline support for online runtime control.}
    \label{tab:offline_support}
    \small
    \setlength{\tabcolsep}{2.8pt}
    \renewcommand{\arraystretch}{1.05}
    \resizebox{\columnwidth}{!}{%
    \begin{tabular}{lll}
    \toprule
    \textbf{Component} & \textbf{Offline output} & \textbf{Runtime role} \\
    \midrule
    Exit interface & Exit module, prediction, metadata & Common execution interface \\
    Latency predictors &  $P_s$ and $P_e$ & Next-exit cost estimation \\
    CNN stabilization & WEL + Mono losses & Stable CNN exit signals \\
    \bottomrule
    \end{tabular}%
}
\end{table}

\sysname separates workload-specific preparation from shared online control. The offline phase exposes candidate exits through a common interface, trains lightweight stage and exit latency predictors from runtime-state profiles, and applies CNN-specific stabilization when needed. This keeps the online control path unified while allowing workload-specific preparation where necessary. Table~\ref{tab:offline_support} summarizes how each offline component supports online runtime.

\subsection{Exit Interface and Model Preparation}
\label{subsec:exit_preparation}

\rev{\sysname supports heterogeneous early-exit workloads through a common interface that exposes ordered stages, executable candidate exits, prediction outputs, and exit-latency metadata. It requires no shared confidence or stability signals, as \sysname’s routing does not use model-specific quality signals.}

For CNN workloads, candidate exits are implemented as branch classifiers attached to selected convolutional or residual blocks. \rev{When exit $k$ is selected, its branch classifier maps $h_k$ to logits $z_k=g_k(h_k)$ and returns the prediction; bypassed classifiers are not executed. Each CNN branch classifier consists of a $1\times1$ convolution, normalization, global average pooling, and a linear classifier. Its architecture is automatically constructed from the feature dimensions at the corresponding exit.}

For Transformer-based language-model workloads, \sysname reuses layer-wise exit heads or pretrained early-exit models. \rev{When an exit layer is selected, its hidden representation is mapped to the output through the model's exit head, while model-specific output semantics and training remain unchanged.}

\rev{\sysname supports models whose exits are trained with confidence-, entropy-, margin-, patience-, or stability-based mechanisms without using these quality signals for routing. Adapting a new architecture only requires mapping its stages and exits to the common interface and profiling their latency; the latency predictors, cache, fallback, and routing rule remain unchanged. Model-specific quality mechanisms can remain as orthogonal components when needed.}

\subsection{Runtime-State-Aware Latency Predictor}
\label{subsec:latency_predictor}

\sysname trains lightweight stage and exit latency predictors offline and uses them online to estimate the remaining latency between candidate exits. Unlike static-profile-based methods \cite{spinn,eenet, e3} that directly use fixed latency tables during deployment, \sysname learns a compact mapping from runtime state to execution latency. Here, runtime state includes the current hardware frequency, resource utilization, and the computation unit being predicted. This design lets the online controller adapt exit decisions when performance changes due to DVFS variation or resource contention.

\textbf{Predictor decomposition.}
\sysname uses two predictors for each workload: a stage-latency predictor $P_s$ and an exit-latency predictor $P_e$. The intuition is that backbone stages and exit modules exhibit different latency patterns. CNN blocks and Transformer layers dominate the sequential execution path, and their latency is affected by computation intensity, memory access, execution depth, and hardware state. Exit modules, such as CNN branch classifiers or language-model exit heads, are auxiliary networks with different computation structures and shorter execution paths. Modeling both with a single predictor may mix heterogeneous latency distributions and reduce prediction accuracy. Therefore, \sysname predicts backbone-stage latency and exit-module latency separately while keeping both predictors compact.

\textbf{Runtime-state features.}
CNN and language-model predictors use six- and five-dimensional feature vectors, respectively. The first three features describe the current frequency state of CPU, GPU, and External Memory Controller (EMC). We normalize each frequency by its maximum supported value:
\begin{equation}
    \hat{f}_{\mathrm{CPU}} = \frac{f_{\mathrm{CPU}}}{f_{\mathrm{CPU}}^{\max}},
    \quad
    \hat{f}_{\mathrm{GPU}} = \frac{f_{\mathrm{GPU}}}{f_{\mathrm{GPU}}^{\max}},
    \quad
    \hat{f}_{\mathrm{EMC}} = \frac{f_{\mathrm{EMC}}}{f_{\mathrm{EMC}}^{\max}}.
    \label{eq:dvfs_normalization}
\end{equation}
The next two features capture runtime load. We use CPU and GPU utilization, denoted as $u_{\mathrm{CPU}}$ and $u_{\mathrm{GPU}}$, to reflect resource contention from co-running tasks. They are normalized as
\begin{equation}
    \hat{u}_{\mathrm{CPU}} = \frac{u_{\mathrm{CPU}}}{100},
    \quad
    \hat{u}_{\mathrm{GPU}} = \frac{u_{\mathrm{GPU}}}{100}.
    \label{eq:utilization_normalization}
\end{equation}
These utilization features allow the predictor to distinguish lightly loaded and contended runtime states even when their DVFS frequencies are identical.

\rev{For CNNs, the final feature is a normalized computation-unit identifier $\mathrm{ID}_\ell$, identifying the block or stage group for stage prediction and the exit module for exit prediction. Language-model predictors omit this feature because Transformer layers are structurally identical.} Overall, the predictor input is
\begin{equation}
    \mathbf{x}_\ell =
    \left[
    \hat{f}_{\mathrm{CPU}},
    \hat{f}_{\mathrm{GPU}},
    \hat{f}_{\mathrm{EMC}},
    \hat{u}_{\mathrm{CPU}},
    \hat{u}_{\mathrm{GPU}},
    \mathrm{ID}_\ell
    \right].
    \label{eq:mlp_input_feature}
\end{equation}

\textbf{Profiling and training.}
During offline profiling, \sysname runs the instrumented model under representative runtime states and records per-stage and per-exit latency. Varying hardware frequencies and utilization levels produces regression samples for the stage and exit predictors. The stage predictor maps $\mathbf{x}_\ell$ to $\hat{L}_\ell$, while the exit predictor maps the corresponding exit feature to $\hat{E}_k$; both are trained with a standard regression loss. Each predictor is a compact two-hidden-layer MLP with 64 and 32 hidden neurons and about 2.5K parameters, keeping latency estimation lightweight on the online decision path. Predictors are calibrated per workload and platform; cross-device transfer is outside this paper's scope.

\textbf{Remaining-latency estimation.}
At runtime, the online controller uses the predictors to estimate whether continuing from the current exit to the next candidate exit is affordable under the remaining latency budget. For candidate exit $k$, \sysname estimates the cost of reaching and evaluating the next candidate exit as
\begin{equation}
    \hat{\Delta L}_{k\rightarrow k+1}
    =
    \sum_{\ell \in \mathcal{P}_{k\rightarrow k+1}}
    \hat{L}_\ell
    +
    \hat{E}_{k+1},
    \label{eq:remaining_latency}
\end{equation}
where $\mathcal{P}_{k\rightarrow k+1}$ denotes the backbone stages between exits $k$ and $k+1$, and $\hat{E}_{k+1}$ is the predicted latency of the next exit module. The online router uses this estimate to decide whether deeper computation is affordable under the current budget.

\subsection{CNN-Specific Exit Stabilization}
\label{subsec:cnn_stabilization}

\sysname uses deeper stages when the latency budget allows, which implicitly assumes that deeper exits provide more reliable predictions. CNN early-exit networks may violate this assumption: due to changes in receptive field and semantic abstraction across depths, a sample that is confident or correct at an early exit may become less confident or even incorrect at a later exit \cite{shallowdeep, multi_exit_curriculum, condmon_earlyexit}. Such non-monotonic behavior can destabilize latency-aware routing. We therefore add a CNN-specific offline stabilization objective, while keeping the online exit interface unchanged. 

\textbf{Weighted exit loss (WEL).}
For the $k$-th CNN exit, the branch classifier produces logits $z_k=g_k(h_k)$ from the intermediate feature $h_k$. Its cross-entropy loss is
\begin{equation}
    \mathcal{L}^{\mathrm{ce}}_k = - \log \frac{\exp(z_{k,y})}{\sum_{c=1}^{C}\exp(z_{k,c})},
    \label{eq:exit_ce_loss}
\end{equation}
where $y$ is the ground-truth label and $C$ is the number of classes. Instead of weighting all exits equally, \sysname assigns larger weights to deeper exits:
\begin{equation}
    \mathcal{L}_{\mathrm{weighted}} = \sum_{k=1}^{K} w_k \cdot \mathcal{L}^{\mathrm{ce}}_k, \quad w_k = w_0 + \alpha \cdot (k-1).
    \label{eq:weighted_exit_loss}
\end{equation}
where $K$ is the number of exits excluding the final model output, $w_0$ is the initial weight, and $\alpha$ controls the linear growth rate. This reduces the dominance of shallow branches over the shared backbone and encourages deeper exits to learn more stable representations.

\textbf{Monotonicity regularization (Mono).}
To reduce confidence degradation across exits, \sysname regularizes the correct-class confidence of adjacent exits. Let $p_k=\mathrm{softmax}(z_k)$ and $p_{k,y}$ be the ground-truth class confidence at exit $k$. The regularization is
\begin{equation}
    \mathcal{L}_{\mathrm{mono}} = \sum_{k=1}^{K-1} \max(0, p_{k,y} - p_{k+1,y}).
    \label{eq:mono_loss}
\end{equation}
This penalty discourages later exits from assigning lower confidence to the correct class than earlier exits, without enforcing strict monotonicity for every sample.

\textbf{Overall objective.}
The final CNN training objective is
\begin{equation}
    \mathcal{L}_{\mathrm{total}} = \mathcal{L}_{\mathrm{final}} + \mathcal{L}_{\mathrm{weighted}} + \lambda_{\mathrm{mono}} \cdot \mathcal{L}_{\mathrm{mono}},
    \label{eq:total_training_loss}
\end{equation}
where $\mathcal{L}_{\mathrm{final}}$ preserves full-model accuracy and $\lambda_{\mathrm{mono}}$ controls the regularization strength. 
This objective stabilizes CNN exits for online routing while preserving the common interface shared with Transformer-based language-model workloads.

\section{Online Runtime Control}
\label{sec:online}

During inference, \sysname follows the online control path in Figure~\ref{fig:uniexit_overview} to make per-stage routing decisions under the current latency budget. At each candidate exit, the controller collects runtime context, estimates next-exit cost, applies cache reuse and fallback as needed, and routes the model to continue or exit.

\subsection{Runtime State Collection}
\label{subsec:runtime_state}

\sysname maintains two lightweight sources of runtime context for online latency prediction and exit routing:
\begin{itemize}[leftmargin=1.2em,itemsep=0.2em,topsep=0.2em]
    \item \rev{\textbf{Hardware-state monitor.} \sysname runs a background thread that periodically samples CPU/GPU/EMC frequencies and CPU/GPU utilization. On Jetson platforms, it obtains CPU frequency via Linux CPUFreq, GPU frequency/utilization via Devfreq, CPU utilization from procfs accounting deltas, and EMC frequency via BPMP. Since some EMC nodes require root access, a privileged process publishes the active EMC frequency to a shared state file. At candidate exits, the inference thread non-blockingly reads the monitor's in-memory snapshot to construct the predictor input in Eq.~\eqref{eq:mlp_input_feature}, avoiding synchronous device queries.}

    \item \rev{\textbf{Inline latency timer.} \sysname measures elapsed latency with a high-resolution monotonic clock, synchronizing CUDA before request execution and at each candidate exit so that preceding GPU kernels complete before timing. The timer is read again after latency prediction to include state retrieval and controller overhead in the remaining-budget calculation. Upon exit, \sysname synchronizes after the exit head and records end-to-end latency. Synchronization is limited to candidate exits rather than every layer to avoid unnecessary GPU serialization.}
\end{itemize}

Together, the monitored state and inline latency measurement provide the context required by the latency predictor and exit router.

\subsection{Latency-Aware Exit Routing}
\label{subsec:exit_routing}

At each candidate exit $k$, \sysname invokes the routing decision before executing the current exit module. Particularly, if the remaining budget still allows deeper execution, \sysname can bypass the current exit module and avoid unnecessary exit overhead. Let $B$ denote the application-specified latency budget, $L_k$ denote the elapsed latency when reaching exit $k$, and $\hat{\Delta L}_{k\rightarrow k+1}$ denote the predicted latency of reaching and evaluating the next candidate exit. \sysname continues to the next stage when $L_k + \hat{\Delta L}_{k\rightarrow k+1} + \Delta S \le B$, and executes the current exit otherwise, where $\Delta S$ is a safety margin used under runtime uncertainty. Therefore, the routing rule to take an action is
\begin{equation}
a_k =
\begin{cases}
\mathrm{Continue}, & L_k + \hat{\Delta L}_{k\rightarrow k+1} + \Delta S \le B,\\
\mathrm{Exit}, & \text{otherwise}.
\end{cases}
\label{eq:decision_rule}
\end{equation}

This rule turns early-exit inference into a deadline-aware runtime control problem. When the predicted next-exit cost fits within the remaining budget, \sysname preserves the opportunity for deeper computation and potentially higher accuracy. When continuing would risk violating the deadline, \sysname executes the current exit module and returns an intermediate result.

\rev{Although \sysname currently focuses on latency-aware routing, its stage-wise control interface could incorporate energy constraints. An additional predictor could estimate the energy cost of reaching the next exit, while the router tracks the remaining energy budget. Under a joint latency--energy policy, execution would continue only if both predicted costs fit their respective budgets; otherwise, it would favor an earlier exit. Such energy-aware prediction and routing are left for future work.}

\begin{figure}[t] 
    \centering
    \includegraphics[width=\linewidth]{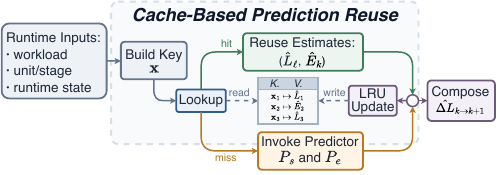}
    \caption{Cache-based prediction reuse in \sysname. The cache is queried before the stage predictor $P_s$ and exit predictor $P_e$; hits reuse recent latency estimates, while misses invoke the corresponding predictor and update the cache.}
    \label{fig:cache_mechanism}
\end{figure}

\subsection{Cache-Based Prediction Reuse}
\label{subsec:cache_reuse}

Although the latency predictors in \sysname are lightweight, online prediction can still become non-negligible when decisions are triggered frequently, especially for token-level language-model exits or workloads with many candidate exits. Meanwhile, edge hardware states often exhibit \textit{temporal locality}: CPU/GPU/EMC frequencies and utilization may remain stable for short intervals, and similar runtime states may recur across consecutive inputs or tokens. This creates an opportunity to reuse recent latency predictions instead of invoking the MLP predictor at every decision point. \rev{The cache does not replace the predictor or act as an offline profiling-based LUT; it is a bounded, on-demand runtime cache storing only recent states and their predictor outputs.}

\textbf{Cache lookup and update.}
As shown in Figure~\ref{fig:cache_mechanism}, \sysname inserts a lightweight cache before both the stage predictor $P_s$ and the exit predictor $P_e$.

\rev{Each entry maps a composite key to a predicted latency. The key combines the workload ID, unit type (stage or exit), and the predictor input in Eq.~\ref{eq:mlp_input_feature}; CNNs use the full $\mathbf{x}_\ell$, while language models omit $\mathrm{ID}_\ell$ because Transformer layers share the same structure. Normalized features are used directly without quantization or bucketization, and hash lookup provides expected $O(1)$ access.}

\rev{Each value is a predicted stage latency $\hat{L}_\ell$ or exit latency $\hat{E}_k$. At candidate exit $k$, cache hits reuse the estimates needed for $\hat{\Delta L}_{k\rightarrow k+1}$ and bypass the MLP; misses invoke $P_s$ or $P_e$ and cache the result.}

\textbf{Bounded reuse.}
To keep the cache small and predictable, \sysname uses a least-recently-used (LRU) replacement policy~\cite{belady1966replacement}. In our implementation, the cache holds 128 entries and occupies about 3~KB; when full, it evicts the least recently accessed entry. This matches the short-term locality of edge runtime states, where recent frequency and utilization configurations are likely to reappear in nearby inference requests. The cache is transparent to model execution: it does not alter the forward path, exit heads, or confidence signals, but only reduces repeated predictor invocations on the online decision path.

\textbf{Interaction with fallback.}
Cache reuse is most effective under static or low-variability runtime states. When the trend-aware module detects volatile hardware behavior, \sysname avoids blindly trusting cached predictions. In such cases, the router either increases the safety margin $\Delta S$ or uses a conservative historical estimate, as described in \S\ref{subsec:fallback}. This keeps cache acceleration from compromising deadline robustness under rapidly changing edge conditions.

\subsection{Trend-Aware Fallback}
\label{subsec:fallback}

Cache reuse and lightweight prediction are effective when runtime states are stable or change gradually. However, abrupt DVFS transitions, thermal throttling, or resource contention can make latency estimates overly optimistic and cause deadline misses. To improve robustness, \sysname uses a trend-aware fallback mechanism.

The fallback module monitors recent runtime-state variations using two indicators: fluctuation magnitude $\sigma_f$ and fluctuation period $\tau_f$. Here, $\sigma_f$ captures the magnitude of recent frequency or utilization changes, while $\tau_f$ captures how rapidly such changes occur. 
When $\sigma_f>\sigma_{\mathrm{thr}}$ and $\tau_f<\tau_{\mathrm{thr}}$, \sysname treats the environment as volatile and augments the safety margin $\Delta S$. The base margin is set from an upper percentile of the offline prediction-error distribution, \eg 95th or 99th percentile. Online, the trend-aware module records periodic worst-case latency and adds the gap between this worst-case latency and the current prediction as extra compensation.

Fallback is triggered only under volatile runtime states and does not replace normal latency prediction. Instead, it complements the predictor by making routing more conservative in tail cases, reducing deadline misses caused by sudden performance drops.

\section{Implementation}
\label{sec:implementation}

We implement \sysname as a runtime layer on top of PyTorch-based CNN and Transformer-based language-model inference pipelines. The runtime includes four main components: \textit{a runtime-state monitor}, \textit{an inline latency timer}, \textit{a cache-assisted latency prediction module}, and \textit{a latency-aware exit router}. The monitor samples CPU/GPU/EMC frequencies and CPU/GPU utilization, and exposes the latest state to the inference thread through shared memory. The inline timer measures elapsed latency at stage and exit boundaries, providing accurate per-request timing for routing decisions.

The latency prediction module uses compact two-hidden-layer MLPs for stage- and exit-latency estimation. Predictors are trained offline from runtime-state profiling data and invoked online only on cache misses. The cache indexes predictions by workload ID, unit type, and normalized predictor features, including stage/exit IDs for CNN workloads, and uses LRU replacement to bound memory usage. The exit router combines elapsed latency, predicted remaining latency, the latency budget, and the fallback safety margin to decide whether to continue or execute the current exit.

For CNN workloads, \sysname inserts branch classifiers at selected convolutional or residual blocks and optionally applies CNN-specific exit stabilization during offline training. For Transformer-based language-model workloads, \sysname reuses layer-wise exit heads or pretrained early-exit variants. This implementation keeps model-specific exits separate from the shared online runtime, allowing the same latency-aware controller to operate across CNN and language-model workloads.

\section{Evaluation}
\label{sec:evaluation}

\begin{table}[t]
    \centering
    \caption{Hardware platforms used in our evaluation.}
    \label{tab:hardware}
    \small
    \setlength{\tabcolsep}{3.5pt}
    \scalebox{0.96}{%
    \begin{tabular}{lccc}
    \toprule
    \textbf{Platform} & \textbf{CPU} & \textbf{GPU} & \textbf{Memory / Compute} \\
    \midrule
    Xavier NX & 6-core Carmel & 384-core Volta & 8 GB / 6.8 TFLOPS  \\
    Orin Nano & 6-core A78AE  & 1024-core Ampere & 8 GB / 10.0 TFLOPS \\
    Cloud Server & 2$\times$32-core Xeon & RTX 5880 Ada & 48 GB / 69.3 TFLOPS \\
    \bottomrule
    \end{tabular}
    }
\end{table}

\subsection{Experimental Setup}
\label{subsec:expsetup}

\textbf{Testbed.}
We evaluate \sysname on Jetson Xavier NX, Jetson Orin Nano, and an edge--cloud platform with an NVIDIA RTX 5880 Ada GPU. Table~\ref{tab:hardware} summarizes their hardware specifications; shared memory and system services reduce usable embedded-device memory by approximately 2~GB from nominal capacity.

\noindent{}\textbf{Workloads and models.}
For CNN workloads, we use ResNet-50~\cite{resnet} and MobileNetV2~\cite{mobilenetv2} to cover both computation-intensive and lightweight perception models. Exit points are inserted at FLOPs-equidistant locations, and ResNet-50 is used as the default CNN backbone unless otherwise noted.

For language-model workloads, we use the 16-layer LLaMA~3.2-1B~\cite{llama32_model_card}, which runs fully on the target platforms without quantization. We focus on this edge-deployable 1B-scale model because larger models require quantization, offloading, or external acceleration, confounding the runtime-control problem. For standalone \sysname, all 16 layers are candidate exits and share the model's original final LM head.

\noindent{}\textbf{Datasets.}
\rev{We use standard datasets for both perception and language-model workloads:}
\begin{itemize}[leftmargin=1.2em,itemsep=0.15em,topsep=0.15em]
    \item \textbf{\textit{CNN inference}.} We use CIFAR-100~\cite{cifar100} and ImageNet~\cite{imagenet}. Training sets are used for supervised training, and test sets are used to generate inference requests. CIFAR-100 images keep their original $32\times32$ resolution, while ImageNet images are resized to $224\times224$. To increase input diversity, we augment 30\% of test samples with random cropping and horizontal flipping.

    \item \rev{\textbf{\textit{Language-model inference}.} We use WikiText-103~\cite{wikitext103}, a standard language-modeling benchmark, for LLaMA~3.2-1B inference. Each input generates a fixed-length output of 64 tokens. We use the full-depth model output on the same inputs as the reference and measure the output consistency of early-exit inference against this reference.}
\end{itemize}

\noindent{}\textbf{Baselines.}
We compare \sysname with representative early-exit architectures and decision methods:
\begin{itemize}[leftmargin=1.2em,itemsep=0.15em,topsep=0.15em]
    \item \textbf{\textit{CNN early-exit architectures}.}
    We compare with SPINN~\cite{spinn} and E3~\cite{e3}, which represent progressive CNN architectures with intermediate branches. For fairness, branch positions and branch counts are aligned with \sysname where applicable.

    \item \textbf{\textit{CNN exit-decision methods}.}
    We compare with EdgeML~\cite{edgeml} and EENet~\cite{eenet}, which optimize early-exit decisions over branch-based CNN networks. These baselines evaluate the benefit of \sysname's latency-aware runtime control beyond conventional confidence-based or static-profile-based decisions.

    \item \textbf{\textit{Transformer-based language-model methods}.}
    We compare with LayerSkip~\cite{layerskip} and SpecEE~\cite{specee} as representative progressive inference methods for language-model workloads. Because \sysname only changes runtime routing and does not modify language-model weights or exit heads, it can be combined with LayerSkip and SpecEE to evaluate orthogonality.

    \item \rev{\textbf{\textit{Concurrent DNN runtime methods}.}
    We compare with Pantheon~\cite{pantheon} and HaX-CoNN~\cite{haxc_onn}, which target concurrent DNN execution under edge GPU and heterogeneous SoC resource contention, respectively. Both baselines evaluate \sysname against contention-aware coordination methods during concurrency.}
\end{itemize}

\begin{figure}[t]
    \centering

    \includegraphics[width=\columnwidth]{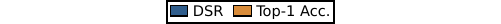}

    \begin{subfigure}[t]{0.495\columnwidth}
        \centering
        \includegraphics[width=\linewidth]{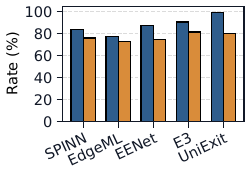}
        \caption{CIFAR-100, Orin Nano}
    \end{subfigure}
    \hfill
    \begin{subfigure}[t]{0.495\columnwidth}
        \centering
        \includegraphics[width=\linewidth]{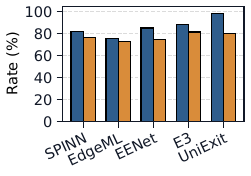}
        \caption{CIFAR-100, Xavier NX}
    \end{subfigure}

    \begin{subfigure}[t]{0.495\columnwidth}
        \centering
        \includegraphics[width=\linewidth]{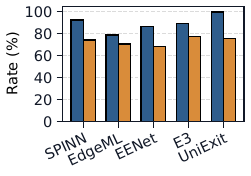}
        \caption{ImageNet, Orin Nano}
    \end{subfigure}
    \hfill
    \begin{subfigure}[t]{0.495\columnwidth}
        \centering
        \includegraphics[width=\linewidth]{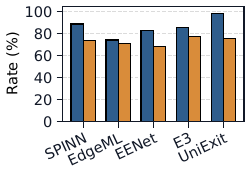}
        \caption{ImageNet, Xavier NX}
    \end{subfigure}

    \caption{Static single-workload performance of CNN early-exit methods across datasets and edge devices.}
    \label{fig:single_static_cnn}
\end{figure}

\noindent{}\textbf{Parameter settings.}
For CNN training, we use SGD with an initial learning rate of 0.1, weight decay of $1\times10^{-4}$, and a batch size of 64. CIFAR learning rates decay at 50\% and 75\% of training epochs, while ImageNet learning rates decay every 30 epochs. The inference batch size is 1 to emulate frame-by-frame edge workloads. \rev{Unless otherwise specified, the relative deadline for each workload is set to its full-depth inference latency under the corresponding hardware setting. Under static maximum-power operation, the deadlines are 15~ms for CNN inference and 70~ms for language-model TPOT; under dynamic DVFS, they are 16~ms and 118~ms, respectively. The same deadline is used across methods within each workload and setting for fair comparison.}
Predictor MLPs use two hidden layers with 64 and 32 neurons. For \sysname-specific parameters, the base safety margin $\Delta S$ is set from an upper percentile of the offline prediction-error distribution and augmented online using the gap between periodic worst-case latency and current prediction. For CNN stabilization, validation search selects $w_0=0.5$ and $\alpha=0.2$, and $\lambda_{\mathrm{mono}}=0.1$. Baseline-specific parameters are tuned on validation settings to maximize their performance.

\noindent{}\textbf{Metrics.}
All results are averaged over five runs. We report:
(i) \textbf{deadline and latency}: \textit{Deadline Satisfaction Rate (DSR)}, \rev{defined as the fraction of requests meeting their deadlines for CNNs and the fraction of inter-token generation intervals meeting their per-token deadlines for language models (TPOT DSR)}, and \textit{latency saving rate};
(ii) \textbf{prediction quality}: \textit{CNN top-1 accuracy} and \textit{language-model relative accuracy}, \rev{defined as the percentage of generated tokens matching the full-depth reference at corresponding positions on the same inputs};
(iii) \textbf{runtime efficiency}: \textit{latency prediction error} and \textit{FLOPs} relative to the full backbone; and
(iv) \textbf{resource cost}: task-level \textit{energy consumption}, computed by integrating device power over the inference interval.

\begin{figure}[t]
    \centering

    \includegraphics[width=\columnwidth]{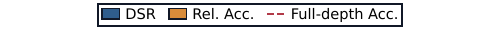}

    \begin{subfigure}[t]{0.495\columnwidth}
        \centering
        \includegraphics[width=\linewidth]{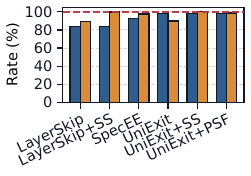}
        \caption{Jetson Orin Nano}
        \label{fig:llm_static_orin}
    \end{subfigure}
    \hfill
    \begin{subfigure}[t]{0.495\columnwidth}
        \centering
        \includegraphics[width=\linewidth]{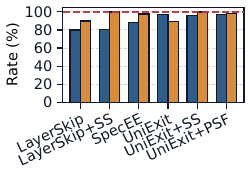}
        \caption{Jetson Xavier NX}
        \label{fig:llm_static_xavier}
    \end{subfigure}

    \caption{\rev{Static single-workload performance of Transformer-based language-model early-exit methods. The dashed line denotes the full-depth reference for relative accuracy.}}
    \label{fig:single_static_llm}
\end{figure}

\subsection{Single-Workload Performance}
\label{subsec:single_workload}

We first evaluate \sysname under isolated single-workload execution for both CNN and Transformer-based language-model workloads. \rev{We consider two controlled settings: (1) \textit{static}, with CPU/GPU/EMC frequencies locked at maximum, and (2) \textit{controlled dynamic DVFS}, where all three frequencies are varied jointly. Orin Nano provides 22/8/4 CPU/GPU/EMC levels and Xavier NX 25/15/6. Starting from maximum frequencies, all dimensions decrease simultaneously by one level every 10~ms; dimensions reaching the minimum early remain there until all reach their minima, after which the sequence reverses analogously to the maxima. This down--up trace repeats throughout execution and is identical across methods, isolating DVFS variation without co-running workloads.}

\textbf{Static hardware states.}
Figure~\ref{fig:single_static_cnn} shows that \sysname achieves the highest and most stable DSR, approaching 100\% across datasets and devices, while outperforming most baselines in accuracy. SPINN, EENet, and E3 also achieve high DSR in some settings but remain sensitive to device- and deadline-dependent latency gaps between adjacent exits; EdgeML is less stable due to online-update overhead. \sysname outperforms SPINN, EdgeML, and EENet in most settings and remains close to E3, which benefits from accuracy-oriented feature enhancement. Overall, combining stable CNN exits with latency-aware routing improves the accuracy--deadline trade-off, showing that even under static hardware, routing must determine whether the next exit is reachable within the remaining budget.

Figure~\ref{fig:single_static_llm} reports the corresponding results for Transformer-based language-model workloads. Standalone \sysname achieves the most stable DSR by making latency-aware routing decisions, but its relative accuracy is lower than accuracy-oriented methods such as LayerSkip and SpecEE, since it does not explicitly model token-level output quality. We further combine \sysname with LayerSkip's self-speculative mechanism (\sysname w/ SS) and SpecEE's probability-shift feature predictor (\sysname w/ PSF). Both combinations improve relative accuracy toward the full-depth reference while preserving high DSR, showing that \sysname is orthogonal to existing accuracy-aware early-exit mechanisms.

\begin{figure}[t]
    \centering
    \includegraphics[width=\columnwidth]{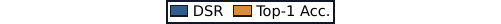}
    
    \begin{subfigure}[t]{0.495\columnwidth}
        \centering
        \includegraphics[width=\linewidth]{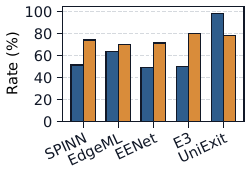}
        \caption{CIFAR-100, Orin Nano}
    \end{subfigure}
    \hfill
    \begin{subfigure}[t]{0.495\columnwidth}
        \centering
        \includegraphics[width=\linewidth]{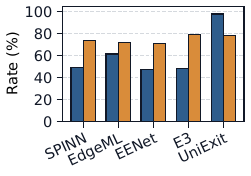}
        \caption{CIFAR-100, Xavier NX}
    \end{subfigure}
    
    \begin{subfigure}[t]{0.495\columnwidth}
        \centering
        \includegraphics[width=\linewidth]{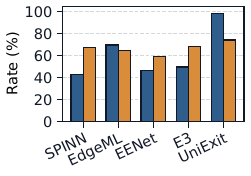}
        \caption{ImageNet, Orin Nano}
    \end{subfigure}
    \hfill
    \begin{subfigure}[t]{0.495\columnwidth}
        \centering
        \includegraphics[width=\linewidth]{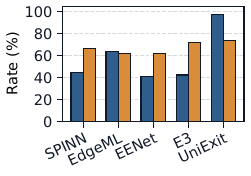}
        \caption{ImageNet, Xavier NX}
    \end{subfigure}
    \caption{CNN early-exit performance under controlled dynamic DVFS across datasets and devices.}
    \label{fig:single_dynamic_cnn}
\end{figure}

\begin{figure}[t]
    \centering

    \includegraphics[width=\columnwidth]{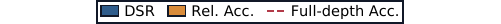}
    \vspace{-1.3em}

    \begin{subfigure}[t]{0.495\columnwidth}
        \centering
        \includegraphics[width=\linewidth]{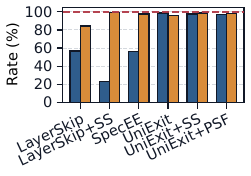}
        \caption{Jetson Orin Nano}
    \end{subfigure}
    \hfill
    \begin{subfigure}[t]{0.495\columnwidth}
        \centering
        \includegraphics[width=\linewidth]{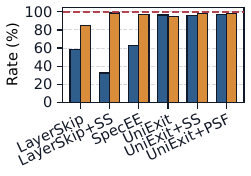}
        \caption{Jetson Xavier NX}
    \end{subfigure}
    \caption{\rev{LLM early-exit performance under controlled dynamic DVFS. Dashed line: full-depth relative accuracy.}}
    \label{fig:single_dynamic_llm}
\end{figure}

\textbf{Controlled dynamic-DVFS states.}
Figure~\ref{fig:single_dynamic_cnn} shows that dynamic DVFS substantially degrades existing CNN early-exit methods. Static-profile-based methods such as SPINN, EENet, and E3 suffer large DSR drops because offline latency tables become stale when hardware frequencies change. EdgeML adapts online, but its updates are too coarse-grained to track fast DVFS variations. In contrast, \sysname maintains high DSR across all CNN settings while preserving strong top-1 accuracy, showing that runtime-state-aware prediction and deadline-aware routing remain effective. The enlarged gap under dynamic DVFS confirms that stale latency profiles are a major source of deadline misses.

Figure~\ref{fig:single_dynamic_llm} reports Transformer-based language-model results under dynamic DVFS; the dashed line marks the full-depth relative-accuracy reference. LayerSkip and SpecEE suffer clear DSR degradation because their routing does not adapt to runtime latency changes. Standalone \sysname maintains substantially higher DSR by adapting to current runtime state and remaining budget, though with lower relative accuracy than accuracy-oriented methods. The \sysname{} w/ SS and \sysname{} w/ PSF variants recover relative accuracy toward the reference while preserving high DSR, confirming latency-safe routing for accuracy-aware early-exit mechanisms.

\begin{figure}[t]
    \centering
    \includegraphics[width=\columnwidth]{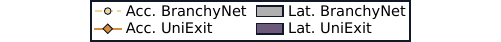}

    \begin{subfigure}[t]{0.495\columnwidth}
        \centering
        \includegraphics[width=\linewidth]{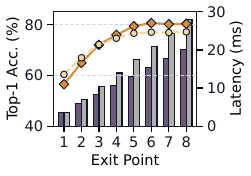}
        \caption{ResNet50, CIFAR-100}
    \end{subfigure}
    \hfill
    \begin{subfigure}[t]{0.495\columnwidth}
        \centering
        \includegraphics[width=\linewidth]{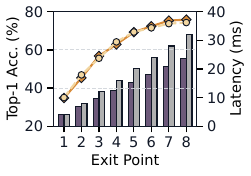}
        \caption{ResNet50, Imagenet}
    \end{subfigure}

    \begin{subfigure}[t]{0.495\columnwidth}
        \centering
        \includegraphics[width=\linewidth]{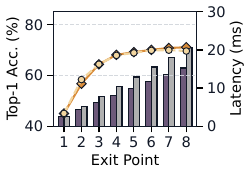}
        \caption{MobileNetV2, CIFAR-100}
    \end{subfigure}
    \hfill
    \begin{subfigure}[t]{0.495\columnwidth}
        \centering
        \includegraphics[width=\linewidth]{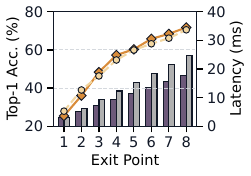}
        \caption{MobileNetV2, Imagenet}
    \end{subfigure}

    \caption{CNN exit-level behavior on Jetson Orin Nano. Lines show top-1 accuracy and bars show cumulative latency.}
    \label{fig:cnn_acc_lat_exit}
\end{figure}

\begin{figure}[!b]
    \vspace*{1.4\baselineskip}
    \centering
    \includegraphics[width=\linewidth]{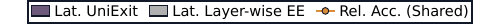}
    \vspace{-1.5em}
    \includegraphics[width=\linewidth]{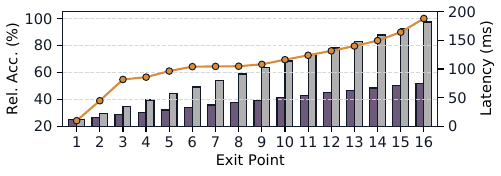}
    \captionsetup{skip=12pt}
    \caption{\rev{Exit-level behavior of the Transformer language model on Orin Nano. Bars show cumulative latency; the line shows \textit{shared} relative accuracy at selected exits.}}
    \label{fig:llm_exit_acc_latency_static}
\end{figure}

\textbf{Exit-level behavior.}
We use the static setting to isolate exit-level behavior, since dynamic DVFS mainly changes latency values rather than selected-exit accuracy.
Figures~\ref{fig:cnn_acc_lat_exit} and~\ref{fig:llm_exit_acc_latency_static} further explain where the latency--accuracy gains of \sysname come from. For CNN workloads, we compare against a BranchyNet-style execution pattern~\cite{branchynet}, which evaluates each encountered branch classifier before deciding whether to exit or continue. As shown in Figure~\ref{fig:cnn_acc_lat_exit}, \sysname consistently reduces cumulative latency across exit points by deciding whether to bypass an exit module before executing it, avoiding repeated branch-evaluation overhead. With two stages between adjacent exits, \sysname advances 1.5$\times$ faster than BranchyNet, further amplifying latency savings at deeper exits. In terms of accuracy, \sysname is slightly lower at a few shallow exits but becomes comparable to or better than BranchyNet at later exits, indicating that CNN-specific exit stabilization improves deeper-exit reliability.

For Transformer-based language-model workloads, we compare against layer-wise EE~\cite{raee} that evaluates intermediate output heads along the layer sequence. As shown in Figure~\ref{fig:llm_exit_acc_latency_static}, \sysname and layer-wise EE share the same relative accuracy at the selected exit layer because they use the same intermediate language-model output. The difference is execution cost: layer-wise EE evaluates intermediate heads along the way, while \sysname skips unselected heads and executes only the selected exit. \sysname thus preserves the same exit-level relative accuracy while substantially reducing cumulative latency, enabling about 2.0$\times$ higher inference progress per unit time than layer-wise EE. These results show that \sysname improves the latency--accuracy trade-off by spending latency budget on backbone progression rather than redundant exit-module evaluation.

\begin{figure}[t]
    \centering

    \includegraphics[width=\columnwidth]{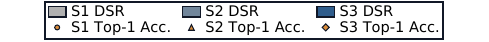}
    \vspace{-1.0em}

    \begin{subfigure}[t]{0.495\columnwidth}
        \centering
        \includegraphics[width=\linewidth]{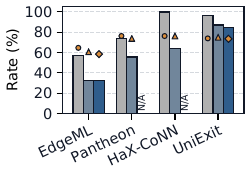}
        \caption{ResNet50, Orin Nano}
    \end{subfigure}
    \hfill
    \begin{subfigure}[t]{0.495\columnwidth}
        \centering
        \includegraphics[width=\linewidth]{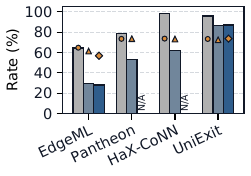}
        \caption{ResNet50, Xavier NX}
    \end{subfigure}

    \begin{subfigure}[t]{0.495\columnwidth}
        \centering
        \includegraphics[width=\linewidth]{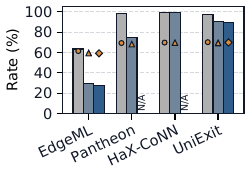}
        \caption{MobileNetV2, Orin Nano}
    \end{subfigure}
    \hfill
    \begin{subfigure}[t]{0.495\columnwidth}
        \centering
        \includegraphics[width=\linewidth]{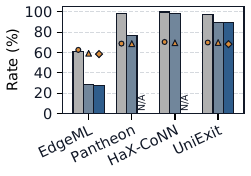}
        \caption{MobileNetV2, Xavier NX}
    \end{subfigure}

    \caption{\rev{CNN performance under heterogeneous concurrent workloads: DSR and Top-1 accuracy. S1: CNN+CNN; S2: CNN+CNN+Map+Audio; S3: CNN+LLM+Map+Audio, with LLaMA~3.2-1B as the LLM.}}
    \label{fig:concurrency_dsr_acc_cnn}
\end{figure}

\begin{figure}[t]
    \centering

    \includegraphics[width=\columnwidth]{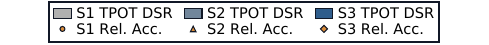}
    \vspace{-1.3em}

    \begin{subfigure}[t]{0.495\columnwidth}
        \centering
        \includegraphics[width=\linewidth]{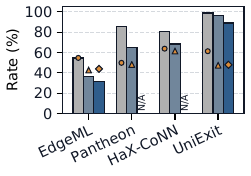}
        \caption{LLaMA-1B, Orin Nano}
    \end{subfigure}
    \hfill
    \begin{subfigure}[t]{0.495\columnwidth}
        \centering
        \includegraphics[width=\linewidth]{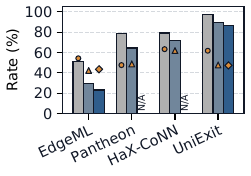}
        \caption{LLaMA-1B, Xavier NX}
    \end{subfigure}

    \caption{\rev{LLM performance under heterogeneous concurrent workloads: TPOT DSR and relative accuracy to full-depth inference. S1: LLM+LLM; S2: LLM+LLM+Map+Audio; S3: CNN+LLM+Map+Audio, with ResNet-50 as the CNN.}}
    \label{fig:concurrency_dsr_acc_llm}
\end{figure}

\begin{figure}[t]
    \centering

    \includegraphics[width=\columnwidth]{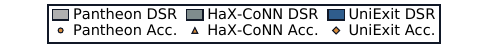}
    \vspace{-1.3em}

    \begin{subfigure}[t]{0.495\columnwidth}
        \centering
        \includegraphics[width=\linewidth]{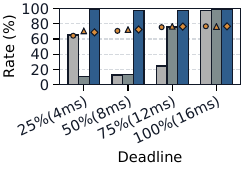}
        \caption{CNN}
        \label{fig:deadline_sensitivity_cnn}
    \end{subfigure}
    \hfill
    \begin{subfigure}[t]{0.495\columnwidth}
        \centering
        \includegraphics[width=\linewidth]{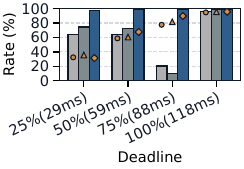}
        \caption{LLM}
        \label{fig:deadline_sensitivity_llm}
    \end{subfigure}

    \caption{\rev{Deadline sensitivity under single-model inference: CNN DSR/Top-1 accuracy at 4--16~ms; LLM TPOT DSR/relative accuracy at 29--118~ms TPOT.}}
    \label{fig:deadline_sensitivity}
\end{figure}

\begin{figure}[t]
    \centering

    \includegraphics[width=\columnwidth]{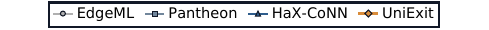}
    \vspace{-1.0em}

    \begin{subfigure}[t]{0.495\columnwidth}
        \centering
        \includegraphics[width=\linewidth]{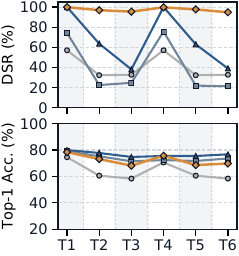}
        \caption{CNN}
        \label{fig:time_varying_cnn}
    \end{subfigure}
    \hfill
    \begin{subfigure}[t]{0.495\columnwidth}
        \centering
        \includegraphics[width=\linewidth]{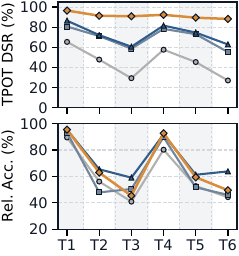}
        \caption{LLM}
        \label{fig:time_varying_llm}
    \end{subfigure}

    \caption{\rev{Performance under time-varying contention over six periods: T1/T4, single model; T2/T5, two models; T3/T6, two models+Map+Audio. CNN reports DSR/Top-1 accuracy and LLM reports TPOT DSR/relative accuracy.}}
    \label{fig:time_varying_dsr_acc}
\end{figure}

\subsection{Concurrent Heterogeneous Workloads}
\label{subsec:concurrent_workloads}

\rev{Under platform-managed DVFS, we evaluate four settings: heterogeneous workloads, time-varying workloads, LUT-based prediction, and deadline sensitivity. Heterogeneous experiments run ResNet-50, MobileNetV2, and LLaMA~3.2-1B on Orin Nano and Xavier NX under three scenarios: two same-type models, two same-type models+Map+Audio, and CNN+LM+Map+Audio. Other settings use ResNet-50 and LLaMA~3.2-1B on Orin Nano. Time-varying execution spans six equal periods repeating single model, two same-type models, and two same-type models+Map+Audio twice. The LUT study replaces the MLP with a DVFS- and utilization-indexed LUT under single-model execution, while deadline sensitivity evaluates methods at 25\%, 50\%, 75\%, and 100\% of the default deadline.}

\rev{Figures~\ref{fig:concurrency_dsr_acc_cnn} and~\ref{fig:concurrency_dsr_acc_llm} show that \sysname maintains high DSR and stable prediction quality under S1. Although heavier CPU/GPU/EMC contention and DVFS variation in S2--S3 reduce DSR, \sysname avoids severe degradation, preserves reasonable prediction quality, and remains robust under heavy heterogeneous workloads.}

\rev{Figure~\ref{fig:deadline_sensitivity} shows that \sysname maintains about 98\% DSR for both workloads as the deadline tightens from 100\% to 25\%, whereas Pantheon and HaX-CoNN degrade substantially and fluctuate more. Looser deadlines enable deeper exits and higher Top-1/relative accuracy, while DSR saturation at 100\% narrows the gap. Overall, \sysname adapts reliably across latency budgets while preserving quality comparable to the baselines.}

\rev{Figure~\ref{fig:time_varying_dsr_acc} shows baseline DSR drops from single-model execution (T1) to two same-type models (T2) and further under Map+Audio contention (T3), recovers at T4, and declines again in T5--T6. In contrast, \sysname maintains high DSR and adapts quickly across transitions while preserving comparable Top-1/relative accuracy.}

\begin{figure}[!t]
    \centering

    \includegraphics[width=\columnwidth]{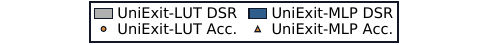}
    \vspace{-1.3em}

    \begin{subfigure}[t]{0.495\columnwidth}
        \centering
        \includegraphics[width=\linewidth]{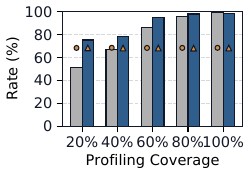}
        \caption{CNN}
        \label{fig:profiling_coverage_cnn}
    \end{subfigure}
    \hfill
    \begin{subfigure}[t]{0.495\columnwidth}
        \centering
        \includegraphics[width=\linewidth]{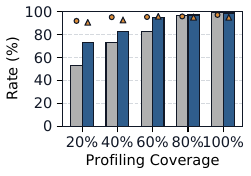}
        \caption{LLM}
        \label{fig:profiling_coverage_llm}
    \end{subfigure}

    \caption{\rev{Profiling-coverage sensitivity of LUT- and MLP-based prediction. CNN reports DSR/Top-1 accuracy; LLM reports TPOT DSR/relative accuracy.}}
    \label{fig:profiling_coverage_sensitivity}

    \par\smallskip

    \includegraphics[width=\columnwidth]{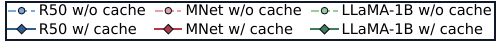}
    \vspace{-1.2em}

    \begin{subfigure}[t]{0.495\columnwidth}
        \centering
        \includegraphics[width=\linewidth]{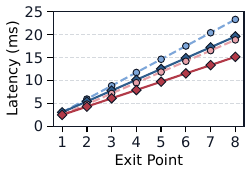}
        \caption{CNN models}
    \end{subfigure}
    \hfill
    \begin{subfigure}[t]{0.495\columnwidth}
        \centering
        \includegraphics[width=\linewidth]{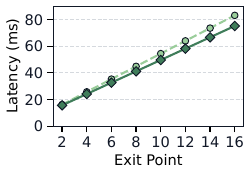}
        \caption{LLaMA3.2-1B}
    \end{subfigure}

    \caption{Latency reduction from cache-based prediction reuse. Enabling cache lowers average inference latency across exit points by avoiding redundant predictor invocations.}
    \label{fig:cache_latency_comparison}

    \par\smallskip

    \includegraphics[width=\linewidth]{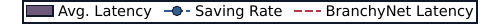}
    \vspace{-1.2em}

    \includegraphics[width=0.99\linewidth]{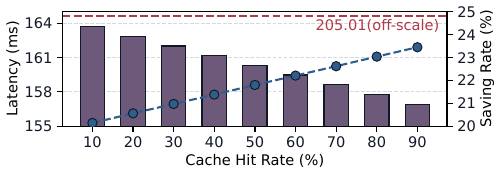}
    \caption{Effect of cache hit rate under dynamic DVFS. Higher cache reuse reduces average inference latency and increases latency saving over branch-based early exit.}
    \label{fig:cache_latency_saving}
\end{figure}

\begin{table}[t]
\centering
\caption{\rev{Memory and runtime overhead of LUT- and MLP-based latency prediction across profiling coverage. \textit{S./I.}: synchronized/independent CPU-frequency scaling; \textit{C./L.}: CNN/LLM. Memory is reported as S./I. and latency as C./L..}}
\label{tab:profiling_mem_lat}

\resizebox{\columnwidth}{!}{
\begin{tabular}{c cc cc}
\hline
\multirow{2}{*}{\textbf{Coverage}} &
\multicolumn{2}{c}{\textbf{LUT-based Prediction}} &
\multicolumn{2}{c}{\textbf{MLP-based Prediction}} \\
\cline{2-5}
&
\textbf{Mem.} &
\textbf{Lat.} &
\textbf{Mem.} &
\textbf{Lat.} \\
&
\textbf{(MiB, S./I.)} &
\textbf{(ms, C./L.)} &
\textbf{(MiB, S./I.)} &
\textbf{(ms, C./L.)} \\
\hline
20\%  & 1.56 / 34.32  & 0.36 / 0.14 & 0.24 / 0.24 & 0.60 / 0.18 \\
40\%  & 3.12 / 68.63  & 0.33 / 0.17 & 0.24 / 0.24 & 0.61 / 0.22 \\
60\%  & 4.68 / 102.95 & 0.29 / 0.15 & 0.24 / 0.24 & 0.61 / 0.23 \\
80\%  & 6.24 / 137.26 & 0.24 / 0.13 & 0.24 / 0.24 & 0.60 / 0.23 \\
100\% & 7.80 / 171.57 & 0.17 / 0.11 & 0.24 / 0.24 & 0.62 / 0.23 \\
\hline
\end{tabular}
}
\end{table}

\rev{Figure~\ref{fig:profiling_coverage_sensitivity} shows that LUT DSR approaches the MLP as profiling coverage increases, whereas the MLP performs better under limited coverage, indicating better generalization to insufficiently profiled states; Top-1 and relative accuracy remain comparable. Table~\ref{tab:profiling_mem_lat} shows that LUT lookup is faster, but its memory reaches 7.80/171.57~MiB at 100\% coverage versus 0.24~MiB for the MLP. Thus, the MLP offers better low-coverage DSR with fixed memory while avoiding exhaustive profiling.}

\rev{Overall, \sysname provides stable latency-aware control across models, devices, contention patterns, and deadlines, while MLP-based prediction improves adaptability to limited profiling and dynamic runtime states.}

\begin{table}[t]
\centering
\caption{Latency prediction error under dynamic DVFS. Each cell reports MAE (ms) / MAPE (\%) / P95 APE (\%). C: CNN; LM: language model; R50: ResNet-50; MNet: MobileNetV2; C100: CIFAR-100; INet: ImageNet; WT103: WikiText-103.}
\label{tab:latency_prediction_error}
\scriptsize
\setlength{\tabcolsep}{2.8pt}
\renewcommand{\arraystretch}{1.03}
\resizebox{\columnwidth}{!}{%
\begin{tabular}{llllcc}
\toprule
\textbf{Type} & \textbf{Model} & \textbf{Data} & \textbf{Dev.} & \textbf{Stage} & \textbf{Exit} \\
\midrule
C  & R50      & C100 & Orin   & 0.112 / 1.08 / 3.42 & 0.086 / 0.94 / 3.05 \\
C  & R50      & C100 & Xavier & 0.147 / 1.31 / 3.86 & 0.113 / 1.17 / 3.48 \\
C  & R50      & INet & Orin   & 0.168 / 1.42 / 4.11 & 0.129 / 0.98 / 3.29 \\
C  & R50      & INet & Xavier & 0.214 / 1.71 / 4.67 & 0.163 / 1.53 / 4.25 \\
\addlinespace[0.15em]
C  & MNet     & C100 & Orin   & 0.094 / 1.14 / 3.36 & 0.071 / 0.88 / 2.91 \\
C  & MNet     & C100 & Xavier & 0.126 / 1.38 / 3.92 & 0.096 / 1.12 / 3.37 \\
C  & MNet     & INet & Orin   & 0.139 / 1.18 / 3.51 & 0.104 / 0.84 / 3.07 \\
C  & MNet     & INet & Xavier & 0.181 / 1.79 / 4.74 & 0.138 / 1.47 / 4.19 \\
\addlinespace[0.15em]
LM & LLaMA-1B & WT103 & Orin   & 0.230 / 1.62 / 5.07 & 0.342 / 1.32 / 4.20 \\
LM & LLaMA-1B & WT103 & Xavier & 0.689 / 2.87 / 6.47 & 0.479 / 1.46 / 4.71 \\
\bottomrule
\end{tabular}%
}
\end{table}

\begin{table}[t]
    \centering
    \caption{\rev{Offline deployment cost on Jetson Orin Nano; training time and FP32 size include both stage and exit predictors.}}
    \label{tab:offline_deployment_cost}
    \small
    \setlength{\tabcolsep}{5pt}
    \renewcommand{\arraystretch}{1.08}
    \begin{tabular}{lcc}
    \toprule
    \textbf{Metric} & \textbf{CNN} & \textbf{LM} \\
    \midrule
    DVFS/utilization states & 17,600 & 17,600 \\
    Profiling records & 52,800 & 52,800 \\
    Profiling time & $\approx$2.73 h & $\approx$29.63 h \\
    Two-MLP training time & $\approx$5.15 min & $\approx$5.14 min \\
    FP32 parameter size & 20.01 KiB & 19.51 KiB \\
    \bottomrule
    \end{tabular}
\end{table}

\subsection{Runtime Efficiency and Overhead}
\label{subsec:runtime_efficiency}

\textbf{Latency prediction accuracy.}
Table~\ref{tab:latency_prediction_error} reports the error of \sysname's runtime-state-aware predictors under dynamic DVFS. Stage and exit predictors are separated because backbone stages and exit modules exhibit different latency distributions. For CNN workloads, MAPE stays below 1.8\% and P95 APE below 4.8\% across models, datasets, and devices. The language-model workload shows higher stage-latency error on Xavier NX due to its more complex DVFS and memory behavior, but the worst MAPE remains below 3.3\%. These errors are low enough for per-exit routing, with tail cases handled by fallback. 

\rev{\noindent\textbf{Offline deployment cost.} Table~\ref{tab:offline_deployment_cost} summarizes the one-time profiling and predictor-training cost on Jetson Orin Nano. These costs are incurred once per workload--platform pair and amortized across inference runs; a new platform requires recalibration but no changes to the shared runtime. CNN exit stabilization is performed only when preparing a new CNN early-exit model.}

\rev{\noindent\textbf{Code reuse and integration cost.} Shared code accounts for 47.2\% of the runtime (52.4\% excluding the platform adapter), while CNN and language-model integration requires only 84 and 152 model-specific SLOC, respectively. Compared with separate runtimes, the shared design reduces logical statements and control-flow branches by 9.3\% and 17.8\%, respectively.}

\textbf{Cache effectiveness.}
Figures~\ref{fig:cache_latency_comparison} and~\ref{fig:cache_latency_saving} evaluate cache-based prediction reuse under static and dynamic settings. Under static hardware, caching reduces average latency across exits, particularly deeper ones with more accumulated predictor calls. Under dynamic DVFS, different change periods yield different hit rates; greater reuse further reduces latency and increases savings over branch-based early exit. Thus, caching remains effective when states are stable or recur over short intervals. Prediction quality remains unaffected because caching skips only redundant predictor calls without altering model execution or exit semantics.

\begin{table}[tb]
    \centering
    \caption{FLOPs ratio relative to the original backbone.}
    \label{tab:flops_ratio}
    \small
    \setlength{\tabcolsep}{4.2pt}
    \renewcommand{\arraystretch}{1.05}
    \begin{tabular}{lccc}
    \toprule
    \textbf{Method} & \textbf{ResNet-50} & \textbf{MobileNetV2} & \textbf{LLaMA~3.2-1B} \\
    \midrule
    Ref. EE  & 133.9\% & 145.7\% & 418.8\% / 248.8\% \\
    \sysname & 101.4\% & 111.6\% & 100.0\% \\
    \bottomrule
    \end{tabular}
    \parbox{0.96\columnwidth}{\footnotesize
    Ref. EE denotes BranchyNet-style CNN and layer-wise language-model early exit. For LLaMA~3.2-1B, 418.8\% and 248.8\% correspond to 16 and 8 exits, respectively.}
\end{table}

\textbf{Computation and energy overhead.}
Table~\ref{tab:flops_ratio} shows that conventional early-exit architectures incur substantial computation by executing intermediate exits: \textit{Ref. EE} increases FLOPs to 133.9\% for ResNet-50, 145.7\% for MobileNetV2, and much higher for LLaMA~3.2-1B due to costly intermediate language-model heads. In contrast, \sysname routes before executing exit modules and skips unselected exits, keeping FLOPs near the original backbone.

\begin{figure}[b]
    \centering
    \begin{subfigure}[t]{0.495\columnwidth}
        \centering        
        \includegraphics[width=\linewidth]{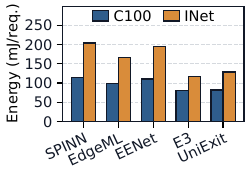}
        \caption{CNN workloads}
    \end{subfigure}
    \hfill
    \begin{subfigure}[t]{0.495\columnwidth}
        \centering
        \includegraphics[width=\linewidth]{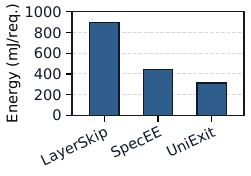}
        \caption{LM workload}
    \end{subfigure}
    \caption{Inference energy on Jetson Orin Nano across CNN and language-model workloads.}
    \label{fig:energy_consumption}
\end{figure}

Figure~\ref{fig:energy_consumption} shows that these computational savings reduce task-level energy consumption. \sysname consumes less energy than branch-heavy CNN baselines and achieves the lowest energy consumption for language-model workloads, confirming that latency-aware routing reduces both computation and energy overhead.

\begin{table}[t]
    \centering
    \caption{Effect of fallback under dynamic DVFS. FB: Fallback; Values are DSR (\%); Gain is the absolute improvement.}
    \label{tab:fallback_ablation}
    \small
    \setlength{\tabcolsep}{4.0pt}
    \renewcommand{\arraystretch}{1.05}
    \begin{tabular}{llccc}
    \toprule
    \textbf{Dataset} & \textbf{Device} & \textbf{w/o FB} & \textbf{w/ FB} & \textbf{Gain} \\
    \midrule
    \multirow{2}{*}{CIFAR-100}
    & Orin   & 94.18 & 98.24 & +4.06 \\
    & Xavier & 93.62 & 97.65 & +4.03 \\
    \addlinespace[0.10em]
    \multirow{2}{*}{ImageNet}
    & Orin   & 94.37 & 98.56 & +4.19 \\
    & Xavier & 92.91 & 97.68 & +4.77 \\
    \addlinespace[0.10em]
    \multirow{2}{*}{WikiText-103}
    & Orin   & 92.82 & 97.41 & +4.59 \\
    & Xavier & 90.72 & 96.46 & +5.74 \\
    \bottomrule
    \end{tabular}
\end{table}

\begin{table}[t]
    \centering
    \caption{Effect of CNN-specific exit stabilization. Values are top-1 accuracy (\%). Equal: equal-weighted exit training; WEL: weighted exit loss; Mono: monotonicity regularization.}
    \label{tab:loss_ablation}
    \small
    \setlength{\tabcolsep}{3.8pt}
    \renewcommand{\arraystretch}{1.05}
    \begin{tabular}{llcccc}
    \toprule
    \textbf{Model} & \textbf{Data} & \textbf{Equal} & \textbf{WEL} & \textbf{Mono} & \textbf{WEL+Mono} \\
    \midrule
    \multirow{2}{*}{ResNet-50}
    & CIFAR-100 & 76.19 & 78.64 & 77.52 & \textbf{80.26} \\
    & ImageNet  & 74.35 & 75.12 & 74.86 & \textbf{75.85} \\
    \addlinespace[0.08em]
    \multirow{2}{*}{MobileNetV2}
    & CIFAR-100 & 71.67 & 72.18 & 71.96 & \textbf{72.59} \\
    & ImageNet  & 68.90 & 69.31 & 69.17 & \textbf{69.68} \\
    \bottomrule
    \end{tabular}
\end{table}

\subsection{Ablation Studies}
\label{subsec:ablation}

\textbf{Effect of fallback.}
Table~\ref{tab:fallback_ablation} shows that fallback consistently improves DSR under dynamic DVFS. Without it, \sysname may underestimate remaining latency when frequencies drop abruptly or cross-exit latency changes faster than the predictor can capture. Fallback increases DSR by 4.03\%--5.74\% using a more conservative safety margin or historical latency estimate. The largest gain occurs for the language-model workload on Xavier NX, which exhibits more variable runtime states. This confirms that fallback handles volatile hardware states rather than replacing latency prediction.

\textbf{Effect of CNN-specific exit stabilization.}
Table~\ref{tab:loss_ablation} evaluates two CNN stabilization losses. \textit{Equal} denotes conventional equal-weighted exit training; \textit{WEL} and \textit{Mono} denote weighted exit loss and monotonicity regularization, respectively. Both improve early-exit accuracy, and their combination performs best across all model--dataset pairs. They are complementary: WEL reduces shallow-exit dominance during shared-backbone training, while Mono stabilizes correct-class confidence across adjacent exits.

\begin{figure}[t]
    \centering
    \includegraphics[width=\columnwidth]{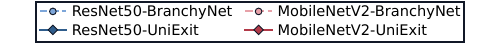}
    \vspace{-1.3em}

    \begin{subfigure}[t]{0.495\columnwidth}
        \centering
        \includegraphics[width=\linewidth]{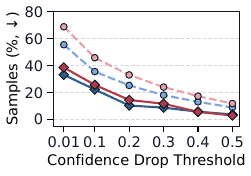}
        \caption{CIFAR-100}
    \end{subfigure}
    \hfill
    \begin{subfigure}[t]{0.495\columnwidth}
        \centering
        \includegraphics[width=\linewidth]{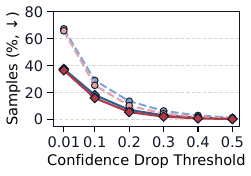}
        \caption{ImageNet}
    \end{subfigure}
    \caption{Confidence degradation across CNN exits. \sysname reduces the fraction of samples whose maximum correct-class confidence drop exceeds each threshold.}
    \label{fig:prob_decrease_samples}
\end{figure}

Figure~\ref{fig:prob_decrease_samples} evaluates exit stability through the fraction of test samples whose maximum correct-class confidence drop exceeds different thresholds. Compared with BranchyNet-style training, \sysname consistently reduces large confidence drops across both datasets and backbones, making deeper CNN exits more reliable when the online router continues within the latency budget.

Together, these ablations show that robust routing requires both runtime-level fallback and model-level exit stabilization.

\begin{figure}[t]
    \centering
    \includegraphics[width=0.94\linewidth]{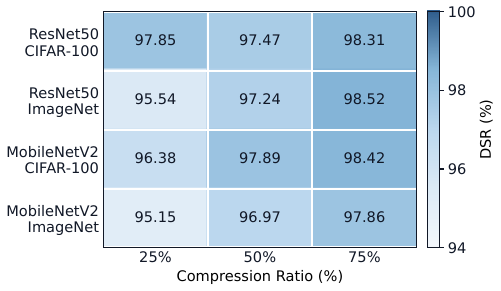}
    \caption{Edge--cloud generalization of \sysname. The heatmap reports DSR under different CNN workloads and intermediate-feature compression ratios.}
    \label{fig:edge_cloud_compression_heatmap}
\end{figure}

\subsection{Generalization to Edge--Cloud Inference}
\label{subsec:edge_cloud}

We assess \sysname's generalization beyond edge-only execution via CNN-based edge--cloud split inference. We focus on CNNs because their feature offloading follows standard split inference, whereas language-model offloading requires transferring hidden states or KV caches and adds scheduling variables~\cite{dualblade}. For simplicity, we split the early-exit CNN midway: Jetson Orin Nano runs the first part and the cloud the remainder. To avoid routing errors from inaccurate transmission estimates, the edge uploads intermediate features, obtains measured transmission feedback, then triggers cloud-side inference with an aligned start signal. \rev{After offloading, the edge continues local inference in parallel and retains the edge-side result as a fallback. If the cloud result is not received before the deadline, \sysname returns the edge-side result as output.}

Figure~\ref{fig:edge_cloud_compression_heatmap} shows that \sysname maintains DSR above 95\% across models, datasets, and feature-compression ratios, with most settings near or above 97\%. Higher compression generally improves DSR by reducing transmission latency and preserving more time for cloud execution. Minor non-monotonic variations suggest that end-to-end timing also depends on model-specific computation and transmission variability. Overall, \sysname remains effective in edge--cloud execution with measured transmission feedback.

\section{Related Work}
\label{sec:related_work}

\textbf{Early-exit inference for CNNs and Transformer-based models.}
Early-exit networks lower average inference cost by exposing intermediate predictions and terminating easy inputs. For CNNs, BranchyNet~\cite{branchynet} attaches side branches, and MSDNet~\cite{msdnet} uses dense connectivity and multi-scale representations for anytime prediction; later work studies branch placement, confidence thresholds, exit training, and accuracy--computation trade-offs~\cite{shallowdeep, early_exit_survey}. For Transformers, DeeBERT~\cite{deebert}, FastBERT~\cite{fastbert}, PABEE~\cite{pabee}, and BERxiT~\cite{berxit} use intermediate classifiers, entropy criteria, patience, or improved fine-tuning. LayerSkip~\cite{layerskip} and SpecEE~\cite{specee} use layer-wise exits with self-speculation or probability-shift features for efficient generation. Although effective model-side mechanisms, their exit heads, confidence signals, and rules remain model-family specific. \rev{Energy-oriented studies adapt execution to energy budgets in energy-harvesting or batteryless settings, including HarvNet~\cite{harvnet}, nonuniformly compressed multi-exit networks for intermittent inference~\cite{intermittent}, and FreeML~\cite{freeml}. In contrast, \sysname extends model-side optimization to latency/deadline-aware, system-level online routing in dynamic edge environments.}

\noindent{}\textbf{Edge-oriented progressive inference systems.}
Progressive systems also target edge or edge--cloud deployment under latency, accuracy, energy, or budget constraints. SPINN~\cite{spinn} combines progressive CNN inference with device--cloud execution and jointly considers early exit and model splitting; EdgeML~\cite{edgeml} adapts policies to changing environments; and EENet~\cite{eenet} and E3~\cite{e3} optimize exit scheduling or progressive architectures for budgeted inference. \rev{Pantheon~\cite{pantheon} and HaX-CoNN~\cite{haxc_onn} study concurrent DNN execution under mobile edge GPU and heterogeneous SoC contention.} Although motivating runtime coordination of early-exit and concurrent inference, many rely on offline profiles, static tables, or periodic policy updates that become inaccurate under DVFS changes, thermal effects, and contention, especially per input or token; most also focus on CNN/DNNs. \sysname instead uses runtime state, inference progress, and latency budgets for online per-exit decisions through one staged interface for CNN and Transformer-based language-model workloads.

\noindent{}\textbf{Latency-aware edge inference and offloading.}
Broader edge inference studies model partitioning~\cite{neurosurgeon, edgent}, offloading~\cite{edgent, deepthings}, latency prediction~\cite{nnmeter, maple_edge}, and hardware-aware adaptation~\cite{dvfo}. Neurosurgeon~\cite{neurosurgeon} partitions DNNs between mobile devices and the cloud, Edgent~\cite{edgent} combines partitioning with model right-sizing, and other work addresses distributed inference, latency modeling, and DVFS-aware adaptation~\cite{deepthings, nnmeter, maple_edge, dvfo}. These systems expose the importance of hardware state, network conditions, and contention, but usually operate per model, request, or partition rather than within the early-exit loop. \rev{They lack runtime-state-aware, deadline-driven control at every candidate exit across heterogeneous model families; \sysname complements them with online runtime-state-aware latency estimation and remaining-budget tracking for per-exit routing in heterogeneous edge AI.}

\section{Conclusion}
\label{sec:conclusion}

We present \sysname, a unified latency-aware early-exit runtime for heterogeneous edge AI workloads. \rev{It treats CNN blocks and Transformer layers as ordered stages with common latency-aware routing while retaining model-specific exit-head design and quality mechanisms.} Runtime-state-aware latency prediction, cache reuse, trend-aware fallback, and CNN-specific exit stabilization enable low-overhead online decisions under dynamic edge conditions. Evaluations on edge and edge--cloud platforms show improved deadline satisfaction and latency--accuracy trade-offs, with less redundant computation and energy overhead, across representative CNN and Transformer-based language-model workloads. \rev{The results validate a shared stage-wise runtime-control abstraction for latency-aware early-exit inference across heterogeneous edge workloads.}

\begin{acks}
We thank our shepherd and the anonymous reviewers for their
constructive feedback.
This work is supported by the National Natural Science Foundation of China
under Grant 62572416, The HKUST-HKUST(GZ) Cross-campus Research Collaboration
``1+1+1'' Joint Funding Program under G\_2025\_052, and Key Areas Project of
the Guangdong Provincial Department of Education under Grant No.~2026ZDZX1022.
The corresponding author is Zhidan Liu.
\end{acks}

\bibliographystyle{ACM-Reference-Format}
\bibliography{references}

\end{document}